\documentstyle[a4,cite,12pt,aps,epsfig]{revtex}

\makeatletter
\def\dgr{\dagger}
\def\nnb{\nonumber}

\def\be{\begin{equation}}
\def\ee{\end{equation}}
\def\mn{\mu\nu}
\def\don{d_1}
\def\dtw{d_2}
\def\dth{d_3}
\newcommand{\bea}{\begin{eqnarray}}
\newcommand{\eea}{\end{eqnarray}}
\makeatother

\begin{document}

\title{The renormalization of the
  effective Lagrangian with
 spontaneous symmetry breaking: the $SU(2)$ case}

\author{
             {\bf Qi-Shu Yan}\footnote{
        E-mail Address: yanqs@mail.ihep.ac.cn} and {\bf Dong-Sheng Du}
\footnote{
        E-mail Address: dsdu@mail.ihep.ac.cn}  \\
        Institute of high energy physics,
	Chinese academy of sciences, Beijing 100039,
	Peoples' Republic of China
}
\bigskip

\address{\hfill{}}

\maketitle

%\begin{center}
%Abstract
%\end{center}
\begin{abstract}
We study the renormalization of the nonlinear effective
$SU(2)$ Lagrangian up to $O(p^4)$ with
spontaneous symmetry breaking.
The Stueckelberg transformation, the background
field gauge, the Schwinger proper time and
heat kernel method, and the
covariant short distance expansion technology,
guarantee the gauge covariance
and incooperate the Ward indentities
in our calculations.
The renormalization group equations of
the effective couplings are derived and analyzed.
We find that the difference between the
results gotten from the direct method and the
renormalization group equation method
can be quite large when the Higgs scalar is
far below its decoupling limit.

\end{abstract}

%\newpage
%\pagestyle{1}
\section{Introduction}

In our last paper \cite{our1}, we discussed the renormalization of
the nonlinear effective $U(1)$ Lagrangian. We learn from the case
that in the framework of effective theory we can do the
renormalization of the nonrenormalizable and
nonlinear interactions order by order. In this paper,
we use those related conceptions and methods to the non-Abelian
cases. We will study the renormalization of the
nonlinear effective $SU(2)$ Lagrangian $L^{eff}$ up to $O(p^4)$
and derive the renormalization group equations (RGE) of its
effective couplings.

We will also numerically study the solutions of these
RGE, and analyze the decoupling and nondecoupling effects of
the Higgs boson to those effective couplings in the
efective Lagrangian $L^{eff}$. We find
that when the Higgs scalr is far below its
decoupling limit, our results are significantly different from
the results gotten by matching the full theory and
effective directly at the one-loop level \cite{dtmd} (Hereby, we call this
method the direct method, in contrast with the RGE method).
The basic reason for this large difference is that the direct method
ignores the contribution of the possible large tree level contributions
of not too heavy Higgs, which can considerably
affect the effective couplings through radiative corrections.
While the RGE method has taken into account these important effects.

The paper is organized as following. In the section II,
we briefly introduce the renormalizable $SU(2)$ Higgs model,
and concentrate on its form in unitary gauge and the quartic divergence
term. In the section III, the nonlinear effective $SU(2)$ Lagrangian
$L^{eff}$ up to $O(p^4)$ is obtained by integrating-out the
scalar Higgs boson at the tree level. We emphasize the importance of the
quartic divergence terms. In the section VI, we perform the renormalization
of the $L^{eff}$ up to $O(p^4)$ in the background field gauge,
and by using the Schwinger proper time and heat kernel
method, derive the renormalization
group equations so as to sum the leading logarithm contributions
of radiative corrections. Section V is devoted to study the
numerical solutions of these RGEs in the Higgs scalar's
decoupling and nondecoupling limits. We end the paper
with some discussions and conclusions.

\section{The renormalizable $SU(2)$ Higgs model}
The partition functional of the renormalizable
non-Abelian $SU(2)$ Higgs model \cite{higgs} (Here we have not included
the gauge fixing term and the ghost term.) can be expressed as
\bea
{\cal Z}=\int {\cal D}A_{\mu}^{a} {\cal D}\phi {\cal D}\phi^{\dgr}
\exp\left (i {{\cal S}[A,\phi,\phi^{\dgr}]}\right )\,,
\eea
where the action ${\cal S}$ is determined by the following Lagrangian
density
\bea
\label{su2l}
{ \cal L} &=& -{1\over 4 g^2} W_{\mn}^a W^{a \mn}
	 + (D\phi)^{\dagger}\cdot(D \phi)
	+\mu^2 \phi^{\dagger} \phi - {\lambda \over 4} (\phi^{\dagger} \phi)^2\,,
\eea
and the definition of quantities in this Lagrangian is given below
\bea
W_{\mn}^a&=&\partial_{\mu} W_{\nu}^a - \partial_{\nu} W_{\mu}^a + f^{abc} W_{\mu}^b W_{\nu}^c\,,\\
D_{\mu} \phi&=&\partial_{\mu} \phi - i W_{\mu}^a T^a \phi\,,\\
\phi^{\dgr}&=&(\phi_1^*,\phi_2^*)\,,
\eea
where $T^a$ are the generators of the Lie algebra of
$SU(2)$ gauge group.

The spontaneous symmetry breaking is induced by
the positive mass square $\mu^2$ in the Higgs potential.
The vacuum expectation value of Higgs
field is given as $|\langle \phi \rangle| = v/{\sqrt 2}$.
And by eating the corresponding
Goldstone boson, the vector bosons $W$ obtain their mass.

The non-linear form of the Lagrangian given in Eqn. (\ref{su2l})
is made by changing the variable $\phi$
\bea
\phi={1\over \sqrt{2}} (v + \rho) U\,,\,
U=\exp\left ({i \xi^a T^a\over v }\right )\,,\,
v=2 \sqrt{\mu^2 \over \lambda}\,,
\eea
where the field $U$ is the Goldstone boson
as prescribed by the Goldstone theorem, and the $\rho$ is a
massive scalar field. Then it reaches
\bea
{\cal L'} &=& -{1\over 4 g^2} W_{\mn}^a W^{a \mn}
	 + {(v+\rho)^2\over 2} (D U)^{\dagger} \cdot (D U)
	+ {1\over 2} \partial \rho \cdot \partial \rho
	+{1\over 2} \mu^2 (v+\rho)^2 - {\lambda \over 16} (v+\rho)^4\,.
\eea
And the change of variables induces a determinant factor in the functional integral ${\cal Z}$
\bea
{\cal Z} = \int {\cal D} W_{\mu}^a {\cal D}\rho {\cal D}\xi^b \exp\left (i {\cal S}'[W,\rho,\xi] \right ) \det\left \{ \left  (1 + {1\over v} \rho\right  ) \delta(x-y) \right \}\,.
\eea
The determinant can be written in the exponential form, and correspondingly the Lagrangian density is modified to
\bea
\label{hml}
{\cal L} \rightarrow {\cal L'} - i \delta(0) ln \left \{ 1 + {1\over v} \rho \right \}\,.
\label{uglag}
\eea
The determinant containing quartic divergences is
indispensable and crucial to cancel exactly the quartic
divergences brought into by the longitudinal part
of vector boson, and is important in verifying the
renormalizability of the Higgs model in the U-gauge \cite{ugauge}.

\section{The nonlinear effective $SU(2)$ Lagrangian $L^{eff}$ up to $O(p^4)$}

In the nonlinear effective $SU(2)$ Lagrangian $L^{eff}$,
only the Goldstone and the vector bosons are included
as the effective dynamic freedom at low energy region.
The Lagrangian $L^{eff}$, if including all permitted
operators composed by these light degrees of freedom (DOFs)
and respecting the assumed Lorentz and gauge symmetries, is
still renormalizable \cite{wein}. Two facts are important for the
actual renormalization procedure:
1) The Wilsonian renormalization method \cite{wilson} and the surface theorem \cite{pol} reveals that
at the low energy region, only few operators play important parts to
determine the behavior of the dynamic system at the low energy region,
such a fact enables us to truncate the infinite divergence
tower and to consider the renormalization
of the effective Lagrangian order by order;
2) While the quartic divergence terms in the effective
Lagrangian enable it possible to consistently
throw away all quartic divergences.

The general effective $SU(2)$ Lagrangian $L^{eff}$ consistent with
Lorenz spacetime symmetry, $SU(2)$ gauge symmetry, and the charge,
parity, and the combined CP symmetries, can be formualted as
\bea
{\cal L}_{eff} &=& {\cal L}_2 + {\cal L}_4 +  \cdots + {\cal L}_{qd}\,,\label{efflag}\\
{\cal L}_2  &=& - {v^2\over 2} tr[V_{\mu} V^{\mu}]\,,\\
{\cal L}_4 &=& -{1\over 4 g^2} W_{\mn}^a W^{a\mn}
- i d_1 tr[W_{\mn} V^{\mu} V^{\nu}]
\nnb\\&&
+ d_2 tr[V_{\mu} V_{\nu}] tr[V^{\mu} V^{\nu}]
+ d_3 tr[V_{\mu} V^{\mu}] tr[V_{\nu} V^{\nu}]
\label{op4}\,,\\
\cdots\,,\nnb\\
{\cal L}_{qd} &=&  i \delta(0) \left \{
 e_1 {tr[V_{\mu} V^{\mu}] \over m_1^2}
+e_2 {(tr[V_{\mu} V^{\mu}])^2 \over m_1^4}
+ \cdots\right  \} \,,
\eea
where the ${\cal L}_2$ and
${\cal L}_4$ represent the relevant and marginal operators in the Wilsonian
renormalization method, respectively. The operators in the ${\cal L}_2$ and
${\cal L}_4$ also form the set of complete operators up to $O(p^4)$ in
the usual momentum counting rule.
And the higher dimension ( irrelevant ) operators than $O(p^4)$ order are represented
by the dots and omitted here. The auxiliary variable $V_{\mu}$ is defined as
\bea
V_{\mu} = U^{\dgr} D_{\mu} U\,,
\eea
to simplify the representation.
Due to the following relations of the $SU(2)$ gauge group
\bea
tr[T^a T^b T^c T^d]={1\over 8} (\delta^{ab} \delta^{cd}+\delta^{ad} \delta^{bc}-\delta^{ac} \delta^{bd})\,,
\eea
the terms, like $tr[V_{\mu} V_{\nu} V^{\mu} V^{\nu}]$ and $tr[V_{\mu} V^{\mu} V_{\nu} V^{\nu}]$,
can be linearly composed by $tr[V_{\mu} V_{\nu}] tr[V^{\mu} V^{\nu}]$
and $tr[V_{\mu} V^{\mu}] tr[V_{\nu} V^{\nu}]$.
And since here we do not consider the term which breaks the charge,
or parity, or both symmetries,
therefore, the operators in Eqn. (\ref{op4}) are complete
and linearly independent.

The effective couplings of $d_i$ form the parameter space of effective
theory, and they effectively reflect the dynamics of the
underlying theories and the ways of symmetry breaking.
Different underlying theories and ways of symmetry breaking
will fall into a special point in this effective parameter space.
And the $d_i$ can also be called the anomalous couplings if seeing from the
renormalizable $SU(2)$ gauge theory, they refect the
deviation from the requirement of renormalizability.

When the scalar Higgs is heavy and integrated out,
the Higgs model given in Eq. (\ref{su2l}) can be effectively
described as a special parameter point of the effective Lagragian
given in Eq. (\ref{efflag}).
At the tree level, it suffices to
integrate out the Higgs scalar boson by using the equation of motion of it,
which expresses it into the low energy dynamic DOFs
and can be formulated as
\bea
\label{eom}
\rho &=& {v \over m_0^2} (D U)^{\dagger}\cdot(D U) + \cdots\,,\\
m_0^2  &=& {1\over 2} \lambda v^2\,,
\eea
where $m_0$ is the mass of Higgs bosons. The omitted
terms contain at least four covariant partials and belong
to higher order operators.

By substituting Eqn. (\ref{eom})
into Eqn. (\ref{hml}), at the matching scale ( which is always taken at the
scalar mass $\mu=m_0$ ) the effective couplings
at the tree level are determined as
\bea
d_1(m_0) = 0\,,\,
d_2(m_0) = 0\,,\,
d_3(m_0) = {v^2 \over 2 m_0^2}={1\over lambda}\,,\,
\cdots\,,
\label{init}
\eea
In its decoupling limit $m_0\rightarrow \infty$, all these three
effective couplings vanish. If a field does not participate in the process of
symmetry breaking, we know it will not contribute to the anomalious couplings
up to the $O(p^4)$ order
and its effects to the low energy dynamics
will be simply suppressed by its squared mass according to the
decoupling theorem \cite{dcpl}.

\section{The renormalization of $L^{eff}$ and its renormalization group equations}

In the background field method (BFM) \cite{bfm, bfm1}, the number of
the Feynman diagrams for the loop corrections can be
greatly decreased when compared with
the standard Feynman diagram method. Another
remarkable advantage is that, in the BFM,
each step of calculation is manifestly gauge
covariant with reference to the background gauge
field, and the Ward identities --- which are important to restrain
the structure of divergences --- have been incorporated in
the calculation. The Schwinger proper time and heat kernel method \cite{htkl} by itself is
the Feynman integral. Combining with the covariant short distance
Taylor expansion \cite{sdte} in coordinate space,
the divergent structures can be directly extracted out
in the explicit gauge form and the loop calculation can be simplified
to a considerable degree.

\subsection{The quadratic terms of the one-loop Lagrangian}

According to the spirit of the BFM, we split
the Goldstone and vector bosons into classic and
quantum parts, as given below
\bea
W\rightarrow {\overline W} + {\widehat W}\,\,,
U\rightarrow {\overline U} {\widehat U}\,,
\eea
The Stueckelberg transformation \cite{sgtn} combines ${\overline W}$
and ${\overline U}$ into the Stueckelberg field ${\overline W}^s$
\bea
{\overline W}^s ={\overline U}^{\dgr} {\overline W} {\overline U}+ i {\overline U}^{\dgr} \partial{\overline U}\,,
\eea
and eliminates the background Goldstone from the effective Lagrangian.
After finishing the loop calculation, by performing the
inverse Stueckelberg transformation (expanding
the ${\overline W^s}$ in the ${\overline W}$
and ${\overline U}$), the effective Lagrangian
can be restored to the form expressed by its low energy DOFs.

As one of the advantages of the BFM, we have the freedom to choose different
gauge for the background and quantum fields, and such a freedom can help to
further simplify the calculation. Then for the quantum fields, we can
choose the covariant gauge fixing term as
\bea
{\cal L}_{GF}=-{1\over 2 g^2}[(D\cdot{\widehat W})^a + c_f f^{abc}{\overline W^{s\,b}} \cdot{\widehat W}^c + f_{ws} \xi^a]^2\,,
\eea
where $c_f$ and $f_{ws}$ are determined by requiring
the one-loop Lagrangian to have the standard form given in
Eqn. (\ref{stdfb}---\ref{stdfe}), then it reads
\bea
c_f = {1\over 2} d_1 g^2\,\,,\,\,f_{ws} = v g^2\,.
\eea

The partition functional ${\cal Z}$ in the background field gauge can be expressed as
\bea
{\cal Z} = \exp\left (i {\cal S}^{ren}[{\overline W^s}]\right ) = \exp\left (i {\cal S}_{tree}[{\overline W^s}] + i \delta {\cal S}_{tree}[{\overline W^s}] + i {\cal S}_{1-loop}[{\overline W^s}] + \cdots\right )\,\nnb\\
=\exp\left (i {\cal S}_{tree}[{\overline W^s}] + i \delta{\cal S}_{tree}[{\overline W^s}]\right ) \int {\cal D}{\widehat W_{\mu}} {\cal D} {\bar c} {\cal D} c {\cal D}\xi \exp\left (i {\cal S}[{\widehat W},\xi,{\bar c},c;{\overline W^s}]\right )\,,
\eea
where the tree effective Lagrangian ${\cal L}_{tree}$ is in the following form
\bea
{\cal L}_{tree}&=& {v^2\over 2} {\overline W^s}\cdot{\overline W^s} -{1\over 4 g^2} {\overline W_{\mn}^{s\,a}} {\overline W^{s\mn,a}}
+ d_1 {1\over 4} f^{abc} {\overline W_{\mn}^{s\,a}} {\overline W^{s\mu,b}} {\overline W^{s\nu,c}}\nnb\\
&&+ d_2 { 1\over 4} {\overline W^{s\,a}} \cdot{\overline W^{s\,b}} {\overline W^{s\,a}}\cdot{\overline W^{s\,b}}
+ d_3 {1\over 4} ({\overline W^s}\cdot{\overline W^s})^2
+ \cdots\,\nnb\\
&&+ i \delta(0) \left [
   e_1 { {\overline W^s} \cdot{\overline W^s} \over m_1^2}
 + e_2 {({\overline W^s}\cdot{\overline W^s})^2 \over m_1^4}
+ \cdots \right  ] \,,
\eea
And the corresponding counter terms $\delta {\cal L}_{tree}$ are defined as
\bea
\delta {\cal L}_{tree} &=&  \delta Z_{v^2} {v^2\over 2} {\overline W^{s\,a}}\cdot{\overline W^{s\,a}}
-\delta Z_{g^2} {1\over 4 g^2} {\overline W_{\mn}^{s\,a}} {\overline W^{s\mn,a}}
+ \delta Z_{d_1} d_1 {1\over 4} f^{abc} {\overline W_{\mn}^{s\,a}} {\overline W^{s\,b\,\mu}} {\overline W^{s\,c\,\nu}}\nnb\\
&&+ \delta Z_{d_2} d_2 { 1\over 4} {\overline W^{s\,a}} \cdot{\overline W^{s\,b}} {\overline W^{s\,a}} \cdot{\overline W^{s\,b}}
+ \delta Z_{d_3} d_3 {1\over 4} ({\overline W^s}\cdot{\overline W^s})^2
+ \cdots\nnb\\
&&+ i \delta(0) \left [
  \delta e_1  { {\overline W^s}\cdot{\overline W^s}\over m_1^2}
+ \delta e_2  {({\overline W^s}\cdot{\overline W^s})^2\over m_1^4}
+ \cdots \right  ] \,,
\label{ctt}
\eea
where the renormalization constant of
the Stueckelberg field ${\overline W^s}$
can always be set to $1$.

In the one-loop level, only the quadratic terms of
quantum fields are related, and they can be
cast into the following standard form
\bea
\label{stdfb}
{\cal L}_{quad}&=&{1\over 2} {\widehat W_{\mu}^a} \Box^{\mu\nu,ab}_{W\,W} {\widehat W_{\nu}^b}
+ {1\over 2} \xi^a  \Box_{\xi\,\xi}^{ab} \xi^b
+ {\bar c}^a  \Box_{{\bar c}c}^{ab} c^b
\nnb\\&&+ {1\over 2} {\widehat W_{\mu}^a} {\stackrel{\leftharpoonup}{X}}^{\mu,ab} \xi^b
+ {1\over 2} \xi^a {\stackrel{\rightharpoonup}{X}}^{\nu,ab} {\widehat W_{\nu}^b}\,,\\
\Box^{\mu\nu,ab}_{W\, W} &=& \left (D'^{2,ab} + m_W^2 \delta^{ab} \right) g^{\mu\nu} - \sigma_{WW}^{\mu\nu,ab}\,\,,\\
\Box_{\xi\,\xi}^{ab} &=& \Box_{\xi\,\xi}'^{ab} + X^{\alpha,ac} d_{\alpha}^{cb} + X^{\alpha\beta,ac} d_{\alpha}^{cd} d_{\beta}^{db}\,\,,\\
\Box_{\xi\,\xi}'^{ab}&=&- \left(d^{2,ab} + \delta^{ab} m_1^2\right ) + \sigma_{2,\xi\xi}^{ab} + \sigma_{4,\xi\xi}^{ab}\,,\\
\Box_{{\bar c}c}^{ab}&=&- \left ( D'^{2,ab} + m_W^2 \delta^{ab} \right)\,,\\
 {\stackrel{\leftharpoonup}{X}}^{\mu,ab}&=&
 {\stackrel{\leftharpoonup}{X}}^{\mu,ac}_{\alpha\beta} d^{\alpha,cd} d^{\beta,db}
+{\stackrel{\leftharpoonup}{X}}^{\mu\alpha,ac} d_{\alpha}^{cb}
+{\stackrel{\leftharpoonup}{X}}^{\mu,ab}_{01}
+{\stackrel{\leftharpoonup}{X}}^{\mu,ab}_{03Z}
+\partial_{\alpha} {\stackrel{\leftharpoonup}{X}}^{\mu\alpha,ab}_{03Y}\,,\\
 {\stackrel{\rightharpoonup}{X}}^{\nu,ab}&=&
 {\stackrel{\rightharpoonup}{X}}^{\nu,ac}_{\alpha\beta} D'^{\alpha,cd} D'^{\beta,db}
+{\stackrel{\rightharpoonup}{X}}^{\nu\alpha,ac} D'^{cb}_{\alpha}
+{\stackrel{\rightharpoonup}{X}}^{\nu,ab}_{01}
+{\stackrel{\rightharpoonup}{X}}^{\nu,ab}_{03Z}
+\partial_{\alpha} {\stackrel{\rightharpoonup}{X}}^{\nu\alpha,ab}_{03Y}\,.
\label{stdfe}
\eea
where $d_{\mu}=\partial_{\mu} - i a_{\xi} {\overline W^s_{\mu,G}}$,
and $D'_{\mu}=\partial_{\mu} - i a_{W} {\overline W^s_{\mu,G}}$.
The direction of the harpoon indicates the position of vector bosons, and
both the ${\stackrel{\leftharpoonup}{X}}^{\mu,ab}$
and ${\stackrel{\rightharpoonup}{X}}^{\nu,ab}$ are defined
to act on the right side.
For the $SU(2)$ effective Lagrangian, the related quantities are defined as
\bea
\label{tbeg}
\sigma^{\mu\nu\,\,ab}_{WW} &=&
 2 i {\overline W^{s\mn,ab}_G} + {1 \over 4} d_1^2 g^4 ({\overline W^{s\mu,ac}_G} {\overline W^{s\nu,cb}_G} - {\overline W}^{s\,ac}_G \cdot {\overline W}^{s\,cb}_G g^{\mn}) \nnb\\
&&-i d_1 g^2 ({\overline W^{s\mn,ab}_G} + {\overline F^{s\mn,ab}_G} )\nnb\\
&&- d_2 g^2 ({\overline W^{s\,a}}\cdot{\overline W^{s\,b}} g^{\mn} + {\overline W^{s\,c}_{\mu}}{\overline W^{s\,c}_{\nu}}\delta^{ab} + {\overline W^{s\,b}_{\mu}}{\overline W^{s\,a}_{\nu}})\nnb\\
&&- d_3 g^2 ({\overline W^{s\,c}}\cdot{\overline W^{s\,c}} g^{\mn} \delta^{ab} + 2 {\overline W^{s\,a}_{\mu}}{\overline W^{s\,b}_{\nu}})\,,\\
\sigma_{2,\xi\xi}^{ab} & = & - a_{\xi}^2 ({\overline W^{s}_G}\cdot{\overline W^{s}_G})^{ab}\,,\\
\sigma_{4,\xi\xi}^{ab} & = & {\widetilde X}_4^{ab}
+ i {1\over 2} a_{\xi} {\widetilde A}_{\alpha \beta}^{ac} ({\overline W^{s \alpha \beta, cb}_G} - {\overline F^{s \alpha \beta, cb}_G})\nnb\\&&
- {\widetilde S}^{\alpha \beta,ac} \Gamma_{\xi,\alpha}^{cd} \Gamma_{\xi, \beta}^{db}
- {\widetilde X^{\alpha,ac}} \Gamma_{\xi,\alpha}^{cb}\,,\\
X^{\alpha,ab} &=& {\widetilde X^{\alpha,ab}}
-  \partial_{\beta} ({\widetilde S}^{\alpha \beta,ab}
+  {\widetilde A}^{\alpha \beta,ab})
+ 2 {\widetilde S}^{\alpha\beta,ac} \Gamma_{\xi,\beta}^{cb}\,,\\
X^{\alpha\beta,ab} &=&- {\widetilde S}^{\alpha \beta,ab}\,,\\
{\stackrel{\leftharpoonup}{X}}^{\mu,ab}_{\alpha\beta}&=&
-{\widetilde S}^{\mu,ab}_{\alpha\beta}\,,\\
{\stackrel{\leftharpoonup}{X}}^{\mu\alpha,ab}&=&
{\widetilde X}_1^{\mu\alpha,ab} - {\widetilde X}_2^{\mu\alpha,ab}
- \partial^{\beta} {\widetilde X}^{\mu,ab}_{\beta\alpha}
+2 {\widetilde S}^{\mu,ac}_{\alpha'\beta} \Gamma^{\beta,cb}_{\xi} g^{\alpha \alpha'}\,,\\
{\stackrel{\leftharpoonup}{X}}^{\mu,ab}_{01}&=&{\widetilde X}^{\mu,ab}_{01}\,,\\
{\stackrel{\leftharpoonup}{X}}^{\mu,ab}_{03Z}&=&
{\widetilde X}^{\mu,ab}_{03}
- {\widetilde S}^{\mu,ac}_{\alpha\beta} \Gamma^{\alpha,cd}_{\xi} \Gamma^{\beta,db}_{\xi}
- ({\widetilde X}_1^{\mu\alpha,ac} - {\widetilde X}_2^{\mu\alpha,ac}) \Gamma^{cb}_{\xi,\alpha}\nnb\\
&&+ i {a_{\xi} \over 2} {\widetilde A}^{\mu,ac}_{\alpha\beta} ({\overline W^{s \alpha \beta, cb}_G} - {\overline F^{s \alpha \beta, cb}_G})\,,\\
{\stackrel{\leftharpoonup}{X}}^{\mu\alpha,ab}_{03Y}&=&- {\widetilde X}_2^{\mu\alpha,ab}\,,\\
{\stackrel{\rightharpoonup}{X}}^{\nu,ab}_{\alpha\beta}&=&
-{\widetilde S}^{\nu,ba}_{\alpha\beta}\,,\\
{\stackrel{\rightharpoonup}{X}}^{\nu\alpha,ab}&=&
{\widetilde X}_2^{\nu\alpha,ba} - {\widetilde X}_1^{\nu\alpha,ba}
-\partial^{\beta} {\widetilde X}^{\mu,ba}_{\alpha\beta}
+2 {\widetilde S}^{\nu,ca}_{\alpha'\beta} \Gamma^{\beta,cb}_{W} g^{\alpha \alpha'}\,,\\
{\stackrel{\rightharpoonup}{X}}^{\nu,ab}_{01}&=&{\widetilde X}^{\nu,ba}_{01}\,,\\
{\stackrel{\rightharpoonup}{X}}^{\nu,ab}_{03Z}&=&
{\widetilde X}^{\mu,ba}_{03}
- {\widetilde S}^{\nu,ca}_{\alpha\beta} \Gamma^{\alpha,cd}_{W} \Gamma^{\beta,db}_{W}
- ({\widetilde X}_2^{\mu\alpha,ca} - {\widetilde X}_1^{\mu\alpha,ca}) \Gamma^{cb}_{W,\alpha}\nnb\\
&&- i {a_{W} \over 2} {\widetilde A}^{\mu,ca}_{\alpha\beta} ({\overline W^{s \alpha \beta, cb}_G} - {\overline F^{s \alpha \beta, cb}_G})\,,\\
{\stackrel{\rightharpoonup}{X}}^{\nu\alpha,ab}_{03Y}&=&
- {\widetilde X}_1^{\nu\alpha,ba}\,,
\eea
where ${\overline F^{s\,a}_{\mn}}=f^{abc} {\overline W^{s\,b}_{\mu}} {\overline W^{s\,c}_{\nu}}$,
${\overline W^{s\,ab}_{\mu,G}}=i f^{acb} {\overline W^{s\,c}_{\mu}}$,
$\Gamma_{\xi,\mu}^{ab}=-i a_{\xi} {\overline W^{s\,ab}_{\mu,G}}$,
and $\Gamma_{W,\mu}^{ab}=-i a_{W} {\overline W^{s\,ab}_{\mu,G}}$, with
$a_{\xi} = 1/2$ and $a_{W} = (1 + d_1 g^2/2)$ (which can be regarded as
the effective charge). To get the above form, we have normalized the
vector quantum gauge field by using ${\widehat W} \rightarrow {\widehat W}/g$.
When we take the limit $d_i \rightarrow 0$,
the $\sigma_{WW}^{\mu\nu,ab}$ reaches to its usual form $2 i W^{\mu\nu,ab}_G$,
as given in the gauge theory without symmetry breaking mechanism.

As in the $U(1)$ case, an auxiliary dimension counting rule is introduced
to extract relevant terms up to $O(p^4)$, which reads
\bea
[{\overline W_{\mu}^s}]_a=[\partial_{\mu}]_a=[D_{\mu}]_a=1\,,\,[v]_a=0\,.
\label{mcnt}
\eea
From this rule, we know
\bea
[{\stackrel{\leftharpoonup}{X}}^{\mu,ab}_{\alpha\beta}]_a=[{\stackrel{\rightharpoonup}{X}}^{\nu,ab}_{\alpha\beta}]_a
=[{\stackrel{\leftharpoonup}{X}}^{\mu,ab}_{01}]_a=[{\stackrel{\rightharpoonup}{X}}^{\nu,ab}_{01}]_a=1\,,
\eea
\bea
[\sigma_{WW}^{\mn\,ab}]_a=[\sigma_{2,\xi\xi}^{ab}]_a=[X^{\alpha\beta,ab}]_a=[{\stackrel{\leftharpoonup}{X}}^{\mu\alpha,ab}]_a
=[{\stackrel{\rightharpoonup}{X}}^{\nu\alpha,ab}]_a=[{\stackrel{\leftharpoonup}{X}}^{\mu\alpha,ab}_{03Y}]_a=[{\stackrel{\rightharpoonup}{X}}^{\nu\alpha,ab}_{03Y}]_a=2\,,
\eea
\bea
[X^{\alpha,ab} ]_a=[{\stackrel{\leftharpoonup}{X}}^{\mu,ab}_{03Z}]_a=[{\stackrel{\rightharpoonup}{X}}^{\nu,ab}_{03Z}]_a=3\,,\,\,
[\sigma_{4,\xi\xi}^{ab}]_a=4\,.
\eea
We would like to mention that this auxiliary dimension counting rule is to extract
those terms with the two, three and four external fields.
In the limit that all anomalous couplings equal to zero,
only the ${\stackrel{\leftharpoonup}{X}}^{\mu,ab}_{01}$,
${\stackrel{\rightharpoonup}{X}}^{\nu,ab}_{01}$,
$\sigma_{WW}^{\mn\,ab}$, and $\sigma_{2,\xi\xi}^{ab}$ do not vanish.

The tilded quantities are determined from the following pre-standard form \cite{bfm1}
\bea
\xi^a \Box_{\xi\,\xi}^{ab} \xi^b&=& - \xi^a \left (d^{2,ab} + \delta^{ab} m_1^2 \right) \xi^b + \xi^a (\sigma_{2,\xi\xi}^{ab} + {\widetilde X}_4^{ab}) \xi^b
\nnb\\&&+ \xi^a {\widetilde X}^{\alpha,ab} \partial_{\alpha} \xi^b+ \partial_{\alpha} \xi^a {\widetilde X}^{\alpha\beta,ab} \partial_{\beta} \xi^b\,\,,\\
{\widehat W^a_{\mu}} {\stackrel{\leftharpoonup}{X}}^{\mu,ab} \xi^b&=& \xi^a {\stackrel{\rightharpoonup}{X}}^{\nu,ab} {\widehat W^b_{\nu}}\nnb\\
&=&\partial^{\alpha} {\widehat W^a_{\mu}} {\widetilde X}^{\mu,ab}_{\alpha\beta}\partial^{\beta} \xi
+ {\widehat W^a_{\mu}} {\widetilde X^{\mu\alpha,ab}_1} \partial_{\alpha} \xi^b
+ \partial_{\alpha} {\widehat W^a_{\mu}} {\widetilde X^{\mu\alpha,ab}_2} \xi^b
\nnb\\&&+ {\widehat W^a_{\mu}} {\widetilde X^{\mu,ab}_{01}} \xi^b
+{\widehat W^a_{\mu}} {\widetilde X^{\mu,ab}_{03}}\xi^b\,.
\eea
and from the effective Lagrangian given in Eqn. (\ref{efflag}), we get
\bea
{\widetilde X}^{\alpha\beta,ab}&=&{\widetilde S}^{\alpha\beta,ab}+{\widetilde A}^{\alpha\beta,ab}\,,\nnb\\
{\widetilde S}^{\alpha\beta,ab}&=&
{d_2\over v^2} \left({\overline W^{s\,a}}\cdot{\overline W^{s\,b}} g^{\alpha\beta}
+ {\overline W^{s\alpha,c}}{\overline W^{s\beta,c}}\delta^{ab}
+ {1\over 2} H^{\alpha\beta,ab}_W\right)\nnb\\
&&+{d_3\over v^2} \left({\overline W^{s\,c}}\cdot{\overline W^{s\,c}} g^{\mn} \delta^{ab}
+H^{\alpha\beta,ab}_W\right)\,,\\
{\widetilde A}^{\alpha\beta,ab}&=& -i {d_1\over 2 v^2} {\overline W^{s\alpha\beta,ab}_G}
+{2 d_3 -d_2\over 2 v^2 } ({\overline W^{s\alpha,a}}{\overline W^{s\beta,b}}
- {\overline W^{s\alpha,b}}{\overline W^{s\beta,a}})\,,\\
{\widetilde X}^{\alpha,ab}&=&
-   {d_1\over 2 v^2} \left({\overline W^{s\alpha\beta,ab}_G} {\overline W^{s,ab}_{\beta,G}}
+{\overline W^{s,ab}_{\beta,G}} {\overline W^{s\alpha\beta,ab}_G}\right )\nnb\\
&&+ i {d_2\over v^2} {\overline W^{s\alpha,c}}{\overline W^{s\beta,c}} {\overline W^{s\,ab}_{\beta,G}}
+ i {d_3\over v^2}  {\overline W^{s\,c}}\cdot{\overline W^{s\,c}} {\overline W^{s\alpha,ab}_{G}}\,,\\
{\widetilde X}^{ab}_4&=&{d_1\over 4 v^2} {\overline W^{s\alpha\beta,ac}_G} {\overline F^{s\,cb}_{\alpha\beta,G}}\,,\\
{\widetilde X}^{\mu,ab}_{\alpha\beta}&=&{\widetilde S}^{\mu,ab}_{\alpha\beta} +{\widetilde A}^{\mu,ab}_{\alpha\beta}\,,\\
{\widetilde S}^{\mu,ab}_{\alpha\beta}&=&
i {d_1 g \over 4 v} (2 {\overline W^{s\mu,ab}_G} g_{\alpha\beta}
- {\overline W^{s\,ab}_{\alpha,G}} g^{\mu}_{\beta}
- {\overline W^{s\,ab}_{\beta,G}} g^{\mu}_{\alpha})\,,\\
{\widetilde A}^{\mu,ab}_{\alpha\beta}&=&- i {d_1 g \over 4 v} ({\overline W^{s\,ab}_{\alpha,G}} g^{\mu}_{\beta}
- {\overline W^{s\,ab}}_{\beta,G} g^{\mu}_{\alpha})\,,\\
{\widetilde X}^{\mu\alpha,ab}_1&=&
    {d_1 g \over 2 v} ({\overline W^{s,ac}_G}\cdot{\overline W^{s,cb}_G} g^{\mu\alpha}
- {\overline W^{s\alpha,ac}_G} {\overline W^{s\mu,cb}_G}
- i {\overline W^{\alpha\mu,ab}_G})\nnb\\
&&- {d_2 g \over   v} ( {\overline W^{s\,a}}\cdot{\overline W^{s\,b}} g^{\mu\alpha}
+ {\overline W^{s\alpha,c}}{\overline W^{s\mu,c}}\delta^{ab}
+ {\overline W^{s\mu,b}}{\overline W^{s\alpha,a}})\nnb\\
&&- {d_3 g \over   v} ( {\overline W^{s\,c}}\cdot{\overline W^{s\,c}} g^{\mu\alpha} \delta^{ab}
+ 2 {\overline W^{s\alpha,b}}{\overline W^{s\mu,a}})\,,\\
{\widetilde X}^{\mu\alpha,ab}_2&=&i {d_1 g\over 2 v} {\overline F^{s\mu\alpha,ab}_G}\,,\\
{\widetilde X}^{\mu,ab}_{01}&=&- i g v (1+ {d_1 g^2 \over 2}) {\overline W^{s\mu,ab}_G}\,,\\
{\widetilde X}^{\mu,ab}_{03}&=&{d_1 g \over 2 v}
 (i{\overline W^{s\mu,ac}_G} {\overline W^{s\alpha,cd}_G} {\overline W^{s\,cb}_{\alpha,G}}
+ i{\overline W^{s\alpha,ac}_G} {\overline W^{s\mu,cd}_G} {\overline W^{s\,cb}_{\alpha,G}}
+  {\overline W^{s\,ac}_{\alpha,G}} {\overline W^{s\mu\alpha,cb}_G})\,,
\label{tend}
\eea
The $H^{\alpha\beta,ab}_W$ is defined as $H^{\alpha\beta,ab}_W={\overline W^{s\alpha,a}}{\overline W^{s\beta,b}} + {\overline W^{s\alpha,b}}{\overline W^{s\beta,a}}$, which
is symmetric on its Lorentz (group) indices.

To get the above form, we have utilized the equation of motion
of the vector bosons ${\overline W^{s}}$, which can be
formulated as
\bea
\partial_{\mu} {\overline W^{s\mn,a}} - {d_1 g^2 \over 2} \partial_{\mu} {\overline F^{s\mn,a}} &=&m_W^2 {\overline W^{s\nu,a}}
- f^{abc} (1-{d_1 g^2 \over 2}){\overline W^{s\,b}_{\mu}}  {\overline W^{s\mn,c}} +{d_1 g^2 \over 2} f^{abc} {\overline W^{s\,b}_{\mu}}  {\overline F^{s\mn,c}}\nnb\\
&&-d_2 g^2 {\overline W^{s\nu,b}} {\overline W^{s\mu,a}} {\overline W^{s\,b}_{\mu}} -d_3 g^2 {\overline W^{s\nu,a}} {\overline W^{s\mu,b}} {\overline W^{s\,b}_{\mu}}\,.
\label{veom}
\eea
From the equation of mition given in Eqn. (\ref{veom}), we can get
\bea
\partial_{\nu} W^{s\nu,a} &=& {1\over m_W^2} \partial_{\nu} \left[
f^{abc} (1-{d_1 g^2 \over 2}) {\overline W^{s\,b}_{\mu}}  {\overline W^{s\mn,c}}
- f^{abc} {d_1 g^2 \over 2} {\overline W^{s\,b}_{\mu}}  {\overline F^{s\mn,c}}
\right.\nnb\\
&&\left.+ d_2 g^2 {\overline W^{s\nu,b}} {\overline W^{s\mu,a}} {\overline W^{s\,b}_{\mu}}
+d_3 g^2 {\overline W^{s\nu,a}} {\overline W^{s\mu,b}} {\overline W^{s\,b}_{\mu}}
\right ]\,.
\eea
Then we know that
the $(\partial_{\mu} W^{s\mu,a})^2$ can only contribute
to terms at most up to $O(p^6)$. Therefore we
simply set $\partial_{\mu} W^{s\mu,a}=0$ when
only considering the renormalization up to $O(p^4)$.

We have also used the following
relations about the Lie algbra:
\bea
f^{abc} f^{cde} + f^{adc} f^{ceb} + f^{aec} f^{cbd} =0\,,
\eea
to simplify the related expressions.

\subsection{The calculation of logarithm and traces}

The quadratic terms can be directly calculated in the functional integral.
Then after integrating out all quantum fields, the ${\cal L}_{1-loop}$ reads
\bea
\int_x {\cal L}_{1-loop}&=&i {1\over 2} \left [Tr\ln\Box_{W\,W}
+Tr\ln\Box_{\xi\,\xi} \right . \nnb\\
& & \left. +Tr\ln\left (1-{\stackrel{\rightharpoonup}{X}^{\mu}} \Box^{-1}_{W\,W;\mu\nu} {\stackrel{\leftharpoonup}{X}^{\mu}} \Box^{-1}_{\xi\,\xi} \right  )\right]
-i Tr\Box_{\bar{c}c}\,\,,
\label{logtr}
\eea
where the contribution of the ghost has a different sign
due to its anticommutator relation.
The $Tr$ is to sum over the Lorentz indices, $\mn$, group indices, $ab$,
and the coordinate space points, $x$. The operators in the term $
Tr\ln\left (1-{\stackrel{\rightharpoonup}{X}^{\mu}} \Box^{-1}_{W\,W;\mu\nu} {\stackrel{\leftharpoonup}{X}^{\mu}} \Box^{-1}_{\xi\,\xi} \right  )$
are all defined to act on the right side, and such a form reflects that
the order of integrating-out the quantum vector boson and Goldstone
fields is unphysical.

The expansion of logarithm is simply expressed as
\bea
\langle x|\ln (1 - X)|y\rangle = - \langle x|X |y\rangle- {1\over 2} \langle x|X X |y\rangle- {1\over 3} \langle x|X X X |y\rangle- {1\over 4} \langle x|X X X X |y\rangle+ ...\,,
\label{tayler}
\eea
and here the $X$ should be understood as an operator (a matrix) which acts on
the quantum states of the right side.

To evaluate the trace, we will use the Schwinger proper time and
heat kernel method \cite{htkl} in the coordinate space.
In this method, the standard propagators can be expressed
as
\bea
\langle x|\Box^{-1,ab}_{W\,W;\mu\nu}|y\rangle=\int_0^{\infty} \frac{d \tau}{(4 \pi \tau)^{d\over 2}} \exp\left(- m_1^2 \tau \right) \exp\left( - {z^2\over 4 \tau}\right) H^{\mu\nu,ab}_{W\,W}(x,y;\tau)\,\,,\label{vpro}\\
\langle x|\Box'^{-1,ab}_{\xi\,\xi}|y \rangle  =\int_0^{\infty} \frac{d \tau}{(4 \pi \tau)^{d \over 2}} \exp\left(- m_1^2 \tau \right) \exp\left( - {z^2\over 4 \tau}\right) H_{\xi\xi,ab}(x,y;\tau)\,\,,
\label{spro}
\eea
where $z=y-x$. The integral over the proper
time $\tau$ and the factor
${\exp\left[ - {z^2/(4 \tau)}\right]/(4 \pi \tau)^{d \over 2}}$
conspire to separate the quadratic divergent part of the propagator.
And the $H(x,y;\tau)$ is analytic with reference to $z$ and
$\tau$, which means that $H(x,y;\tau)$ can be analytically
expanded with reference to both $z$ and $\tau$. Then we have
\bea
H(x,y;\tau)&=&H_0(x,y) +H_1(x,y) \tau + H_2(x,y) \tau^2 + \cdots\,,
\eea
where $H_0(x,y)$, $H_1(x,y)$, and, $H_2(x,y)$ are
the Silly-De Witt coefficients. The coefficient
$H_0(x,y)$ is the pure
Wilson phase factor, which indicates the
phase change of a quantum state
when moving from the point $x$ to the point $y$ and
reads
\bea
H_0(x,y)= C \exp \left(- \int_y^x \Gamma(z)\cdot dz\right ),
\eea
where $\Gamma(z)$ is the affine connection ( dependent on
the group representation of the quantum states ) defined on the
coordinate point $z$. And the coefficient $C$ is related with
the spin of the states, for vector bosons, $C=-g_{\mu\nu}$,
and for scalar bosons, $C=1$.

The divergence counting rule of the integral
over the coordinate space $x$ and the proper
time $\tau$ can be established as
\bea
[z^{\mu}]_d=1\,\,,\,\,[\tau]_d=-2\,.
\label{dcnt}
\eea

Using Eqn. (\ref{tayler}), the two propagators
defined in Eqn. (\ref{vpro}) and Eqn. (\ref{spro}),
and the divergence and momentum counting rule given
in Eqn. (\ref{mcnt}) and (\ref{dcnt}),
up to $O(p^4)$, we can get the following divergent terms
\bea
{\bar \epsilon} Tr\ln\Box_{WW}&=&\int_x \left [-m_W^2 tr[g_{\mu\nu} \sigma_{WW}^{\mu\nu}]
+ {8\over 3} ({1\over 4} {\Gamma^{a}_{W,\mn}} {\Gamma^{\mn,a}_W})
\right.\nnb\\&&\left.+ {1\over 2} tr[\sigma_{WW}^{\mu\mu'} (-g_{\mu'\nu'}) \sigma_{WW}^{\nu'\nu} (-g_{\mu\nu})] \right ]\,,\label{vctt}\\
{\bar \epsilon} Tr\ln\Box_{{\bar c}c}&=&\int_x \left [
 {2\over 3} ({1\over 4} {\Gamma^{a}_{W,\mn}} {\Gamma^{\mn,a}_{W}})
 \right ]\,,\label{gstt}\\
{\bar \epsilon} Tr\ln\Box_{\xi\xi}&=& \int_x \left [m_W^2 tr[\sigma_{2,\xi\xi}]
+ m_W^2 tr[\sigma_{4,\xi\xi}]
+ {2\over 3} ({1\over 4} {\Gamma^{a}_{\xi,\mn}} {\Gamma^{\mn,a}_{\xi}})\right.\nnb\\
&&\left.+ {1\over 2} tr[\sigma_{2,\xi\xi} \sigma_{2,\xi\xi}]
+ {1\over 2} m_W^2 tr[X^{\alpha\beta} \sigma_{2,\xi\xi}]\right ]\,,\label{sclt}\\
Tr\ln\left (1-{\stackrel{\rightharpoonup}{X}^{\mu}} \Box^{-1}_{W\,W;\mu\nu} {\stackrel{\leftharpoonup}{X}^{\mu}} \Box^{-1}_{\xi\,\xi} \right )&=&
{1\over {\bar \epsilon}}  (p4t+p3t+p2t)\,,
\eea
where $1/{\bar \epsilon}=i (2/\epsilon - \gamma_E + \ln4 \pi^2)/(16 \pi^2)$, $\gamma_E$ is
the Euler constant, and $\epsilon=4-d$. The
$\Gamma_{\mn}$ is the field strength
tensor corresponding to the affine connection
$\Gamma_{\mu}$.
We have used the dimensional regularization scheme
and the modified minimal substraction scheme to
extract the divergent structures in this step.
The $p4t$ represents the
contributions of four propagators $tr(
{\stackrel{\rightharpoonup}{X}} \Box^{-1}_{W\,W} {\stackrel{\leftharpoonup}{X}} \Box'^{-1}_{\xi\,\xi}
{\stackrel{\rightharpoonup}{X}} \Box^{-1}_{W\,W} {\stackrel{\leftharpoonup}{X}} \Box'^{-1}_{\xi\,\xi}
)$, which reads
\bea
p4t&=&g_{\mu\nu} g_{\mu'\nu'} \left[ {g^{\alpha\beta\alpha'\beta'}\over 4}
 tr[2 {\stackrel{\rightharpoonup}{X}}^{\mu}_{\alpha\beta} {\stackrel{\leftharpoonup}{X}}^{\nu}_{\alpha'\beta'} {\stackrel{\rightharpoonup}{X}}^{\mu'}_{01} {\stackrel{\leftharpoonup}{X}}^{\nu'}_{01} \right. \nnb\\
&&\left. + 2 {\stackrel{\rightharpoonup}{X}}^{\mu}_{01} {\stackrel{\leftharpoonup}{X}}^{\nu}_{\alpha\beta} {\stackrel{\rightharpoonup}{X}}^{\mu'}_{\alpha'\beta'} {\stackrel{\leftharpoonup}{X}}^{\nu'}_{01}
+   {\stackrel{\rightharpoonup}{X}}^{\mu}_{\alpha\beta} {\stackrel{\leftharpoonup}{X}}^{\nu}_{01} {\stackrel{\rightharpoonup}{X}}^{\mu'}_{\alpha'\beta'} {\stackrel{\leftharpoonup}{X}}^{\nu'}_{01}
+   {\stackrel{\rightharpoonup}{X}}^{\mu}_{01} {\stackrel{\leftharpoonup}{X}}^{\nu}_{\alpha\beta} {\stackrel{\rightharpoonup}{X}}^{\mu'}_{01} {\stackrel{\leftharpoonup}{X}}^{\nu'}_{\alpha'\beta'}]\right.\nnb\\
&&\left.+m_W^2 {g^{\alpha\beta\alpha'\beta'\alpha''\beta''}\over 4}
 tr[{\stackrel{\rightharpoonup}{X}}^{\mu}_{\alpha\beta} {\stackrel{\leftharpoonup}{X}}^{\nu}_{\alpha'\beta'} {\stackrel{\rightharpoonup}{X}}^{\mu'}_{\alpha''\beta''} {\stackrel{\leftharpoonup}{X}}^{\nu'}_{01}
+ {\stackrel{\rightharpoonup}{X}}^{\mu}_{\alpha\beta} {\stackrel{\leftharpoonup}{X}}^{\nu}_{\alpha'\beta'} {\stackrel{\rightharpoonup}{X}}^{\mu'}_{01} {\stackrel{\leftharpoonup}{X}}^{\nu'}_{\alpha''\beta''}]
\right ]\,.
\eea
The $p3t$ represents the contributions of three propagators $tr(
{\stackrel{\rightharpoonup}{X}} \Box^{-1}_{W\,W} {\stackrel{\leftharpoonup}{X}} \Box'^{-1}_{\xi\,\xi} X_{\alpha\beta} d^{\alpha} d^{\beta} \Box'^{-1}_{\xi\,\xi})$, which reads
\bea
p3t&=&{m_W^2 \over 4} g^{\alpha\beta\alpha'\beta'} g_{\mn}
tr[{\stackrel{\rightharpoonup}{X}}^{\mu}_{01} {\stackrel{\leftharpoonup}{X}}^{\nu}_{\alpha\beta} X_{\alpha'\beta'}
+ {\stackrel{\rightharpoonup}{X}}^{\mu}_{\alpha\beta} {\stackrel{\leftharpoonup}{X}}^{\nu}_{01} X_{\alpha'\beta'}]
+{1\over 2} g^{\alpha\beta} g_{\mn} tr[{\stackrel{\rightharpoonup}{X}}^{\mu}_{01} {\stackrel{\leftharpoonup}{X}}^{\nu}_{01} X_{\alpha\beta}]\,.
\eea
The $p2t$ represents the contributions of two propagrators $tr(
{\stackrel{\rightharpoonup}{X}} \Box^{-1}_{W\,W} {\stackrel{\leftharpoonup}{X}} \Box'^{-1}_{\xi\,\xi})$, which can be further
divided into six groups:
\bea
p2t&=&t_{AA} + t_{AB} + t_{AC} + t_{BB} + t_{BC} + t_{CC}\,,\\
t_{AA}&=&{m_W^2 g^{\mn}\over 4} ({g^{\alpha\beta\alpha'\beta'\delta\gamma} \over 6} - { g^{\alpha\beta} g^{\alpha'\beta'\delta\gamma} \over 2}
-{g^{\alpha'\beta'} g^{\alpha\beta\delta\gamma} \over 2}+ g^{\alpha\beta} g^{\alpha'\beta'} g^{\delta\gamma})
tr[{\stackrel{\rightharpoonup}{X}}^{\mu}_{\alpha\beta} D_{\delta}D_{\gamma}{\stackrel{\leftharpoonup}{X}}^{\nu}_{\alpha'\beta'}]\nnb\\
&&+{m_W^2  g^{\alpha\beta\alpha'\beta'} \over 8} \left [
g_{\mn} tr[{\stackrel{\rightharpoonup}{X}}^{\mu}_{\alpha\beta}{\stackrel{\leftharpoonup}{X}}^{\nu}_{\alpha'\beta'} \sigma_{2,\xi\xi}]
- g_{\mu\mu'} g_{\nu\nu'} tr[{\stackrel{\rightharpoonup}{X}}^{\mu}_{\alpha\beta}\sigma^{\mu'\nu'}_{WW}{\stackrel{\leftharpoonup}{X}}^{\nu}_{\alpha'\beta'}]\right ]\,,\nnb\\
t_{AB}&=&{m_W^2  g_{\alpha'\alpha''} g_{\mu\nu} \over 2} (g^{\alpha\beta}g^{\alpha'\beta'} -{1\over 2} g^{\alpha\beta\alpha'\beta'} )
tr[{\stackrel{\rightharpoonup}{X}}^{\mu}_{\alpha\beta} D_{\beta'} {\stackrel{\leftharpoonup}{X}}^{\nu\alpha''}
- {\stackrel{\rightharpoonup}{X}}^{\mu\alpha''} D_{\beta'} {\stackrel{\leftharpoonup}{X}}^{\nu}_{\alpha\beta}]\,,\\
t_{AC}&=& - {m_W^2  g^{\alpha\beta} g_{\mn} \over 2}
tr[{\stackrel{\rightharpoonup}{X}}^{\mu}_{\alpha\beta} {\stackrel{\leftharpoonup}{X}}^{\nu}_{01}
+{\stackrel{\rightharpoonup}{X}}^{\mu}_{01} {\stackrel{\leftharpoonup}{X}}^{\nu}_{\alpha\beta}
+{\stackrel{\rightharpoonup}{X}}^{\mu}_{\alpha\beta} {\stackrel{\leftharpoonup}{X}}^{\nu}_{03Z}
+{\stackrel{\rightharpoonup}{X}}^{\mu}_{03Z} {\stackrel{\leftharpoonup}{X}}^{\nu}_{\alpha\beta}\nnb\\
&&-\partial_{\alpha'} {\stackrel{\rightharpoonup}{X}}^{\mu}_{\alpha\beta} {\stackrel{\leftharpoonup}{X}}^{\nu\alpha'}_{03Y}
-{\stackrel{\rightharpoonup}{X}}^{\mu\alpha'}_{03Y} \partial_{\alpha'} {\stackrel{\leftharpoonup}{X}}^{\nu}_{\alpha\beta}]\nnb\\
&&-{1\over 4} g^{\alpha\beta}
 tr[g_{\mu\mu'} g_{\nu\nu'} {\stackrel{\rightharpoonup}{X}}^{\mu}_{\alpha\beta} \sigma^{\mn'\nu'}_{WW} {\stackrel{\leftharpoonup}{X}}^{\nu}_{01}
-g_{\mu\nu} {\stackrel{\rightharpoonup}{X}}^{\mu}_{01} {\stackrel{\leftharpoonup}{X}}^{\nu}_{\alpha\beta} \sigma_{2,\xi\xi}]\,\nnb\\
&&-g_{\mu\nu} ({1\over 6} g^{\alpha\beta\alpha'\beta'} - {1\over 4} g^{\alpha\beta} g^{\alpha'\beta'})
tr [{\stackrel{\rightharpoonup}{X}}^{\mu}_{\alpha\beta} D_{\alpha'}D_{\beta'}{\stackrel{\leftharpoonup}{X}}^{\nu}_{01}
+{\stackrel{\rightharpoonup}{X}}^{\mu}_{01} D_{\alpha'}D_{\beta'}{\stackrel{\leftharpoonup}{X}}^{\nu}_{\alpha\beta}]\,,\\
t_{BB}&=&-{m_W^2 g_{\mn} g_{\alpha\beta} \over 2} tr[{\stackrel{\rightharpoonup}{X}}^{\mu\alpha} {\stackrel{\leftharpoonup}{X}}^{\nu\beta}]\,,\\
t_{BC}&=&- { g^{\alpha\beta} g_{\alpha\alpha'} g_{\mn} \over 2}
tr[{\stackrel{\rightharpoonup}{X}}^{\mu\alpha'} D_{\beta} {\stackrel{\leftharpoonup}{X}}^{\nu}_{01}
-{\stackrel{\rightharpoonup}{X}}^{\mu}_{01} D_{\beta} {\stackrel{\leftharpoonup}{X}}^{\nu\alpha'}]\,,\\
t_{CC}&=&-g_{\mn} tr[{\stackrel{\rightharpoonup}{X}}^{\mu}_{01} {\stackrel{\leftharpoonup}{X}}^{\nu}_{01}
+{\stackrel{\rightharpoonup}{X}}^{\mu}_{01} {\stackrel{\leftharpoonup}{X}}^{\nu}_{03Z}
+{\stackrel{\rightharpoonup}{X}}^{\mu}_{03Z} {\stackrel{\leftharpoonup}{X}}^{\nu}_{01}
-\partial_{\alpha'} {\stackrel{\rightharpoonup}{X}}^{\mu}_{01} {\stackrel{\leftharpoonup}{X}}^{\nu\alpha'}_{03Y}
-{\stackrel{\rightharpoonup}{X}}^{\mu\alpha'}_{03Y} \partial_{\alpha'} {\stackrel{\leftharpoonup}{X}}^{\nu}_{01}]\,.
\label{mixt}
\eea
where the trace is to sum over the group indices and points
of coordinate space, and
the covariant differentials is defined as
\bea
{\stackrel{\rightharpoonup}{X}} D {\stackrel{\leftharpoonup}{X}} &=& {\stackrel{\rightharpoonup}{X}} \partial {\stackrel{\leftharpoonup}{X}}
+ {\stackrel{\rightharpoonup}{X}} \Gamma_W {\stackrel{\leftharpoonup}{X}}
- {\stackrel{\rightharpoonup}{X}} {\stackrel{\leftharpoonup}{X}} \Gamma_{\xi}\,,\\
{\stackrel{\rightharpoonup}{X}} D D {\stackrel{\leftharpoonup}{X}} &=& {\stackrel{\rightharpoonup}{X}} \partial \partial {\stackrel{\leftharpoonup}{X}}
+ {\stackrel{\rightharpoonup}{X}} \Gamma_W \Gamma_W {\stackrel{\leftharpoonup}{X}}
+ {\stackrel{\rightharpoonup}{X}} {\stackrel{\leftharpoonup}{X}} \Gamma_{\xi} \Gamma_{\xi}
-2 {\stackrel{\rightharpoonup}{X}} \Gamma_W {\stackrel{\leftharpoonup}{X}} \Gamma_{\xi}
\nnb\\&&+ 2 {\stackrel{\rightharpoonup}{X}} \Gamma_W \partial {\stackrel{\leftharpoonup}{X}}
- 2 {\stackrel{\rightharpoonup}{X}} \partial {\stackrel{\leftharpoonup}{X}} \Gamma_{\xi}
+  {\stackrel{\rightharpoonup}{X}} \partial \Gamma_W {\stackrel{\leftharpoonup}{X}}
-  {\stackrel{\rightharpoonup}{X}} {\stackrel{\leftharpoonup}{X}} \partial \Gamma_{\xi}\,.
\eea
And the tensors $g^{\alpha\beta\gamma\delta}$ and $g^{\alpha\beta\gamma\delta\mu\nu}$
are symmetric on all indices and defined as
\bea
 g^{\alpha\beta\gamma\delta} &=& g^{\alpha\beta} g^{\gamma\delta} + g^{\alpha\gamma} g^{\beta\delta} + g^{\alpha\delta} g^{\beta\gamma}\,,\\
 g^{\alpha\beta\gamma\delta\mu\nu} &=& g^{\alpha\beta} g^{\gamma\delta\mu\nu} + g^{\alpha\gamma} g^{\beta\delta\mu\nu} + g^{\alpha\delta} g^{\gamma\beta\mu\nu}
+g^{\alpha \mu} g^{\beta\gamma\delta\nu} + g^{\alpha\nu} g^{\beta\gamma\delta\mu}\,.
\eea
To get the $p4t$, $p3t$, and $p2t$, we have used the covariant
shord-distance expansion technology \cite{sdte} and the integral over
the proper time and coordinate space. We would like to comment on the
the covariant shord-distance expansion technology: to formulate
the quadratic form into the standard form can simplify the labor
to extract the divergences, while the form given in \cite{sdte}
is not easy to use. The equivalence of these two forms
can be easily proved by using the partial integral. As we have pointed
out, the standard form given by us has the advantage to reflect the
fact that the order of integrating out the quantum vector boson and
Goldstone fields has no any dynamic significance, and is unphysical.

\subsection{The renormalization group equations}
Substituting Eqs. (\ref{tbeg}---\ref{tend}) to Eqs. (\ref{vctt}---\ref{mixt}), with
somewhat tedious algebraic manipulation, we construct the counter terms
and extract the renormalization constants. The renormalization constants
yield the following RGEs
\bea
\label{rgeb}
{d g^2\over dt} &=&{g^4\over 8 \pi^2} \left[
-\frac{29}{4}   + 6 \don g^2 -
  \frac{17 {\don}^2 g^4}{8}
\right]\,,\\
{d v\over dt}   &=&{ v \over 16\pi^2} \left [ -
\frac{3 g^2}{2} - \left( 5 \don + 8 \dtw +
     14 \dth \right)  g^4 - \frac{7 {\don}^2 g^6}{2}
 \right ]\,,\\
{d d_1\over dt} &=&{ 1 \over 8 \pi^2} \left \{
- \frac{1}{12}  +
  \left( \frac{11 \don}{2} - \frac{11 \dtw}{2} +
     11 \dth \right)  g^2  \right.\nnb\\&&\left.+
  \left[ - \frac{69 {\don}^2}{8} +
     \don \left( 3 \dtw - 6 \dth \right)
\right]  g^4 + \frac{23 {\don}^3 g^6}{48}
 \right \}\,,\\
{d d_2 \over dt}&=&{ 1 \over 8 \pi^2} \left \{
- \frac{1}{12}  +
  \left( -\frac{119 \don}{48} + 6 \dtw \right)  g^2 +
  \left( \frac{173 {\don}^2}{24} +
     20 \don \dtw \right)  g^4 \right.\nnb\\&&\left.+
  \left[ \frac{207 {\don}^3}{16} +
     {\don}^2 \left( \frac{91 \dtw}{8} -
        \frac{5 \dth}{4} \right)  \right]  g^6 +
  \frac{545 {\don}^4 g^8}{96}
 \right\}\,,\\
{d d_3\over dt} &=&{ 1 \over 8 \pi^2} \left\{
- \frac{1}{24}  +
  \left( \frac{89 \don}{48} - \frac{19 \dtw}{2} -
     11 \dth \right)  g^2 \right.\nnb\\&&\left.+
  \left[ \frac{251 {\don}^2}{48} -
     \don \left( 28 \dtw +
        \frac{61 \dth}{2} \right)  \right]  g^4 \right.\nnb\\&&\left.+
  \left[ \frac{239 {\don}^3}{16} -
     {\don}^2 \left( \frac{53 \dtw}{4} +
        \frac{43 \dth}{4} \right)  \right]  g^6 +
  \frac{463 {\don}^4 g^8}{96}
 \right \}\,.
\label{rgee}
\eea
About the RGEs given in
Eqs. (\ref{rgeb}---\ref{rgee}),
it is remarkable that the direct
method will only get part of
the result of the RGE method,
which is contributed by the Goldstone boson and
indicated by the constant
terms independent of $d_i,\,i=1,\,2,\,3$
in the rhs of RGEs of  $d_i,\,i=1,\,2,\,3$.
While the rest terms of the RGEs take into account not only
the effect of Goldstone boson $\xi$, but also that
of vector bosons ${\widehat W}$ and that of their mixing terms.

In order to compare and contrast, we formulate
the results of the direct method in the RGE form,
which read
\bea
\label{rgeb0}
{d g^2\over dt} &=&{g^4\over 8 \pi^2} \left[- \frac{29}{4}\right]\,,\\
{d v\over dt}   &=&{ v \over 16\pi^2} \left [-\frac{3 \, g^2}{2}  \right ]\,,\\
{d d_1\over dt} &=&{ 1 \over 8 \pi^2} \left [ - \frac{1}{12}  \right ]\,,\\
{d d_2 \over dt}&=&{ 1 \over 8 \pi^2} \left [ -\frac{1}{12} \right]\,,\\
{d d_3\over dt} &=&{ 1 \over 8 \pi^2} \left [-\frac{1}{24}  \right ]\,.
\label{rgee0}
\eea
The underlying reason to represent the contributions
of Higgs in the RGE form might be related with the fact that
the full theory is a renormalizable one and the divergences
generated by the Higgs loop should be cancelled out exactly by
those generated by the Goldstone bosons. To extract the divergent
structures, we have used the following relations of the $SU(2)$ gauge
group
\bea
tr[W_G^{s\mu} W_G^{s\nu} W^s_{\mu,G} W^s_{\nu,G}] &=& 2 {\overline W^{s\,a}} \cdot{\overline W^{s\,b}} {\overline W^{s\,a}}\cdot{\overline W^{s\,b}}\,,\\
tr[W_G^{s\mu} W^{s}_{\mu,G} W^{s\nu}_G W^s_{\nu,G}] &=& {\overline W^{s\,a}} \cdot{\overline W^{s\,b}} {\overline W^{s\,a}}\cdot{\overline W^{s\,b}} + ({\overline W^s}\cdot{\overline W^s})^2 \,,\\
H_{\mu\nu}^a H^{\mu\nu,a} &=& W_{\mn}^a W^{a \mn} - 2 f^{abc} {\overline W_{\mn}^{s\,a}} {\overline W^{s\mu,b}} {\overline W^{s\nu,c}} \nnb\\&&+
({\overline W^s}\cdot{\overline W^s})^2 - {\overline W^{s\,a}} \cdot{\overline W^{s\,b}} {\overline W^{s\,a}}\cdot{\overline W^{s\,b}}\,,
\eea
where the variable $H_{\mu\nu}^a$ is symmetric when exchanging its
Lorentz indices $\mu$ and $\nu$, and is defined as
$H_{\mu\nu}^a = \partial_{\mu} W^{s\,a}_{\nu} + \partial_{\nu} W^{s\,a}_{\mu}$.

The term $11 d_3 g^2$ in the right hand side of the
RGE of $d_1$ is quite remarkable: the coeffecient
$11$ mainly comes from the fact that the gauge bosons
have $3$ physical components as a vector field, and have
$3$ components as an adjoint representation of the
$SU(2)$. When the Higgs is not too heavy, the
coupling $d_3$ can reach order $0.1$ or $0.01$,
then this term can switch the sign of $d_1(m_W)$
from positive to negative.
Another remarkable fact is that it is the radiative corrections
from the vector bosons which mostly contribute to the
linear terms of $d_i$ in the rhs of RGEs and dominate the
running of RGEs when the nonlinear effects of the RGEs
is still small.
In the following section we will comparatively study the
solutions of these two groups of RGEs.

\section{Numerical analysis}

We concentrate on the Higgs scalar's effects to the
effective couplings $d_i,\,i=1,\,2,\,3$. To simplify the
analysis, we mimic the standard model by choosing the
mass of vector boson $m_W$ to be $91$ GeV. The Higgs
scalar is assumed to be heavier than the vector bosons
$W$. The initial condition for the coupling $g$ and
the vacuum expectation value $v$ is fixed at the
lower boundary point, $\mu=m_W$.
The coupling $g(m_W)$ is chosen to satisfy
\bea
\alpha_g={g^2\over 4\pi}={1\over 30}\,,
\eea
which gives $g(m_W)=0.65$.
and the vacuum expectation value is then
fixed by $m_W=g v$, which gives $v(m_W)=140.6$.
While the initial condition for $d_i,\,i=1,\,2,\,3$
is chosen to be fixed at the matching scale, $\mu=m_0$,
as given in Eqn (\ref{init}).

Below we will compare the results gotten from
the direct method and the RGE method.
As we know, the scalar's effect includes
both the decoupling mass square suppressed
part as shown in Eqn. (\ref{init})
and the nondecoupling logarithm part.
So we consider the following three cases to
trace the change of roles of these two competing
parts:
1) the light scalar case, with $m_0 = 160$ GeV,
where the decoupling mass square suppressed
part dominates;
2) the not too heavy scalar case, with $m_0 = 500$ GeV,
where both contributions are important;
3) the very massive scalar case, with $m_0 = 1.2$ TeV,
where the nondecoupling logarithm part dominates.

The figure 1. is devoted to the first case. It is
obvious that the difference of $d_i(m_W),\,i=1,\,2$ in
two methods is quite large, and the $d_1(m_W)$'s
have different signs in these two method.
Figure 2. is devoted to the second case. For the $d_1$,
these two methods predict different sign with the same magnitudes.
Figure 3. is devoted to the third case. And the difference between
the results of these two methods is relatively small.

In the all three cases, the magnitude of
the $d_2(m_W)$ is about $10^{-3}$ in both methods.
and the difference between these two methods is neglectable.

\begin{figure}
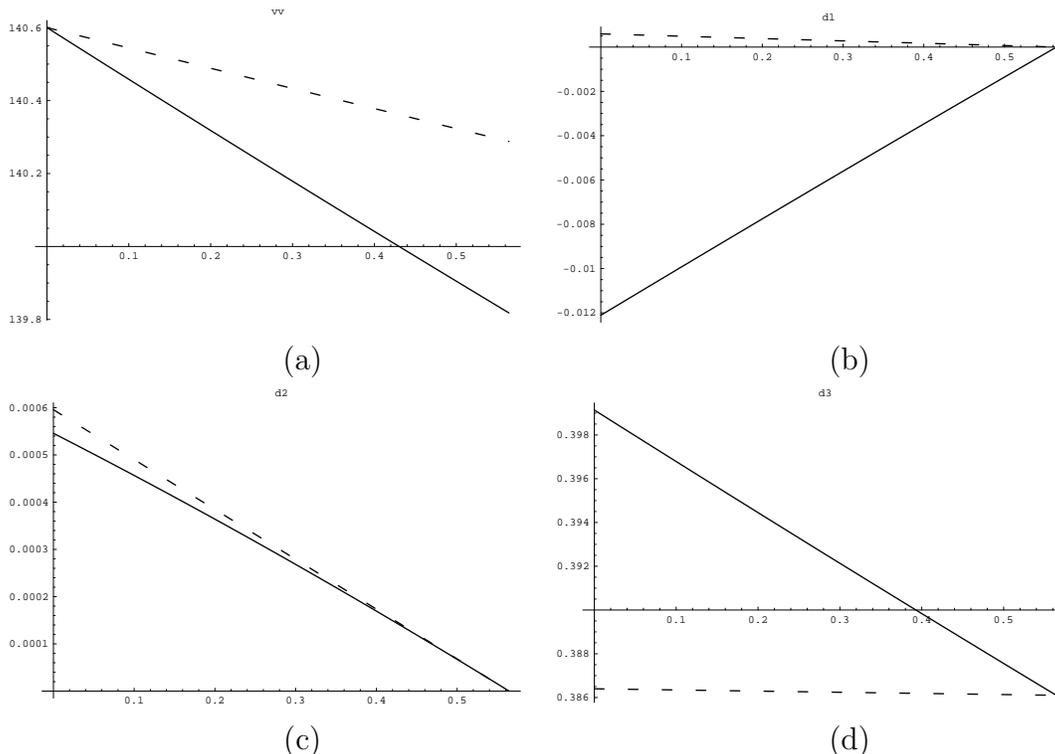

\begin{minipage}[t]{6.8cm}
     \epsfig{file=tmp11.epsi,width=6.8cm}
     \mbox{ }\hfill\hspace{1cm}(a)\hfill\mbox{ }
     \end{minipage}
     \hspace{0.2cm}
     \begin{minipage}[t]{6.8cm}
     \epsfig{file=tmp12.epsi,width=6.8cm}
     \mbox{ }\hfill\hspace{1cm}(b)\hfill\mbox{ }
     \end{minipage}
\vskip 0.05truein
     \begin{minipage}[t]{6.8cm}
     \epsfig{file=tmp13.epsi,width=6.8cm}
     \mbox{ }\hfill\hspace{1cm}(c)\hfill\mbox{ }
     \end{minipage}
     \hspace{0.2cm}
     \begin{minipage}[t]{6.8cm}
     \epsfig{file=tmp14.epsi,width=6.8cm}
     \mbox{ }\hfill\hspace{1cm}(d)\hfill\mbox{ }
     \end{minipage}
     \caption{\it
The varying of $v$, $d_1$, $d_2$, and $d_3$ with the
running scale $t$ ($t=\ln(m_0/m_W)$).
The matching scale is the mass of Higgs
scalar, which is taken to be $m_0=160$ GeV.
The solid lines are the results of the RGE method,
while the dashed ones are those of the direct method.}
\label{fig1}
\end{figure}

While the $d_1(m_W)$ can reach $10^{-2}$ in the RGE method,
one order larger than in the direct method,
when the Higgs scalar is quite small.
Even when the Higgs is medium heavy, the results
of these two method have the same magnitude and different
signs. Near the decoupling limit, the prediction
of RGE method improves that of the direct method
up to $40\%$---$70\%$.

Due to its initial values at the matching scale,
the $d_3(m_W)$ could have different magnitudes in these
three cases, $10^{-1}$, $10^{-2}$, and $10^{-3}$, respectively.
The differences of these two methods are small when the Higgs is
relative light, and in the third case the relative difference
can reach $5\%$---$15\%$.

The difference of the running of $g$ is neglectable in these two
methods so we have not depicted it.

From these figures, we can read out the
tendency that the difference of $\delta d_1$ between
these two methods is larger when the Higgs scalar
is further below its decoupling limit.
The underlying reason for this behavior is
related with the initial value $d_3$ at the matching scale,
and the related terms dependent on $d_3$
in the RGEs given in Eq. (\ref{rgeb}---\ref{rgee}).

\begin{figure}
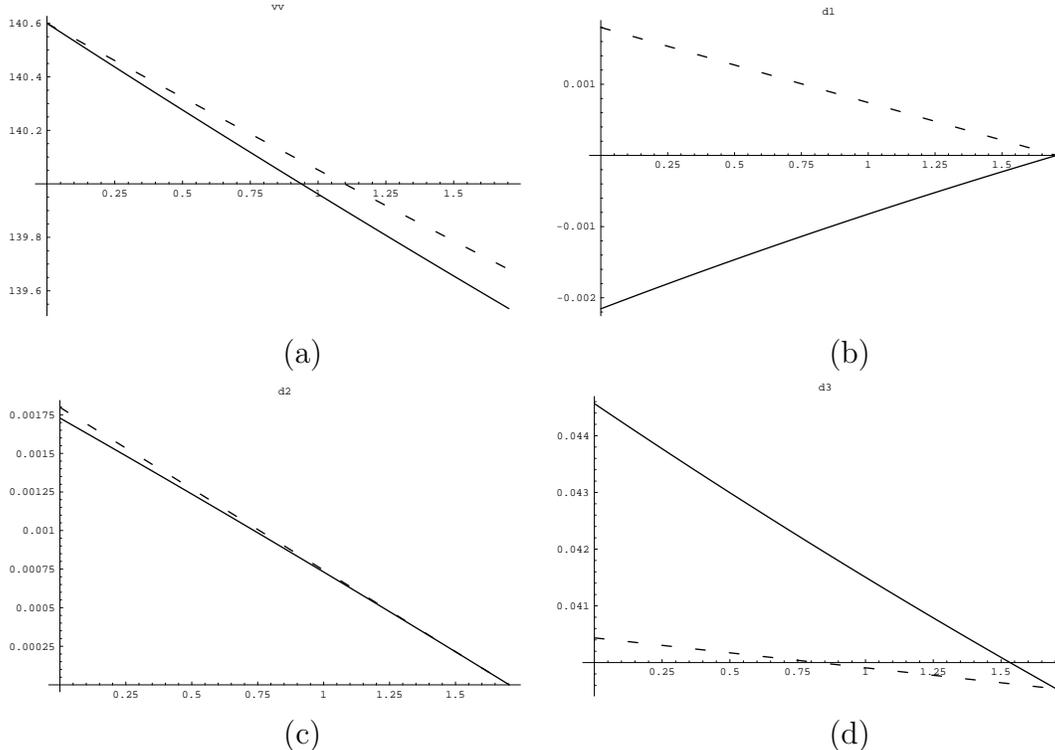

\begin{minipage}[t]{6.8cm}
     \epsfig{file=tmp21.epsi,width=6.8cm}
     \mbox{ }\hfill\hspace{1cm}(a)\hfill\mbox{ }
     \end{minipage}
     \hspace{0.2cm}
     \begin{minipage}[t]{6.8cm}
     \epsfig{file=tmp22.epsi,width=6.8cm}
     \mbox{ }\hfill\hspace{1cm}(b)\hfill\mbox{ }
     \end{minipage}
\vskip 0.05truein
     \begin{minipage}[t]{6.8cm}
     \epsfig{file=tmp23.epsi,width=6.8cm}
     \mbox{ }\hfill\hspace{1cm}(c)\hfill\mbox{ }
     \end{minipage}
     \hspace{0.2cm}
     \begin{minipage}[t]{6.8cm}
     \epsfig{file=tmp24.epsi,width=6.8cm}
     \mbox{ }\hfill\hspace{1cm}(d)\hfill\mbox{ }
     \end{minipage}
     \caption{\it
The varying of $v$, $d_1$, $d_2$, and $d_3$ with the
running scale $t$ ($t=\ln(m_0/m_W)$). The matching scale is the mass of Higgs
scalar, which is taken to be $m_0=500$ GeV.
The solid lines are the results of the RGE method,
while the dashed ones are those of the direct method.}
\label{fig2}
\end{figure}

\section{Discussions and Conclusions}
In this paper, we have studied the renormalization of the nonlinear
effective $SU(2)$ Lagrangian with spontaneous symmetry breaking and
derived its RGEs. Compared with the $U(1)$ case,
the non-Abelian case is much more complicated.
And quite differently, in the $SU(2)$ case, the gauge coupling
and the anomalous couplings up to $O(p^4)$ are driven
to develop by the quantum fluctuations low energy DOFs.
We also have comparatively studied the results
of the direct method and the RGE method.
From the numerical analysis, we see that
the results of two methods are very
different when the Higgs scalar is far below its
decoupling limit. The underlying reason is
related with the initial value of $d_3$ at the
matching scale and with the radiative correction
of all low energy degrees of freedom ( both the
Goldstone and vector bosons ) which contributes
to the $d_3$ terms in Eq. (\ref{rgeb}---\ref{rgee}).

Normally, when the Higgs is so light, the higher
dimension operators, for instance those belong to
the $O(p^6)$ order, might play some considerable
parts and it might be not good to use the
effective theory to describe the full theory, since
the Wilsonian renormalization \cite{wilson} and the surface theorem given
by \cite{pol} require that the low energy scale $\mu_{IR}$ is
lower enough than the UV cutoff.
While we see, for the medium heavy Higgs (say,
$m_0 = 400$ GeV or $m_0 = 800$ GeV ), it is still approperiate to
use it, and the difference between these two method is quite considerable
for some anomalous coupling(s).

\begin{figure}
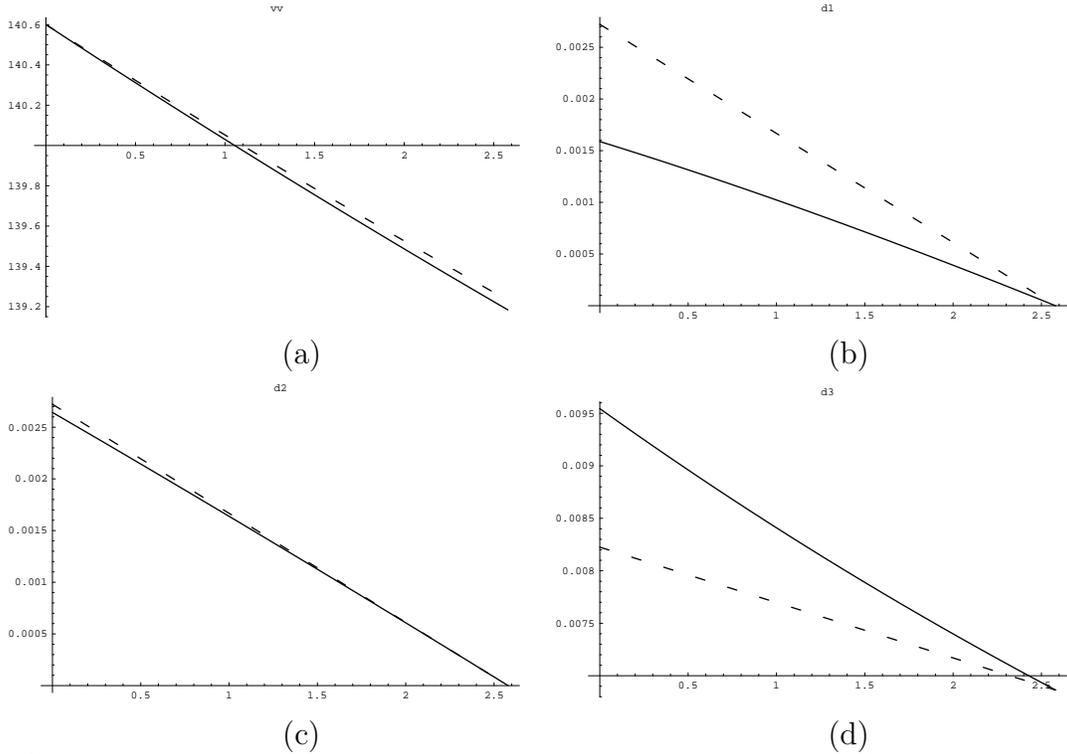

\begin{minipage}[t]{6.8cm}
     \epsfig{file=tmp31.epsi,width=6.8cm}
     \mbox{ }\hfill\hspace{1cm}(a)\hfill\mbox{ }
     \end{minipage}
     \hspace{0.2cm}
     \begin{minipage}[t]{6.8cm}
     \epsfig{file=tmp32.epsi,width=6.8cm}
     \mbox{ }\hfill\hspace{1cm}(b)\hfill\mbox{ }
     \end{minipage}
\vskip 0.05truein
     \begin{minipage}[t]{6.8cm}
     \epsfig{file=tmp33.epsi,width=6.8cm}
     \mbox{ }\hfill\hspace{1cm}(c)\hfill\mbox{ }
     \end{minipage}
     \hspace{0.2cm}
     \begin{minipage}[t]{6.8cm}
     \epsfig{file=tmp34.epsi,width=6.8cm}
     \mbox{ }\hfill\hspace{1cm}(d)\hfill\mbox{ }
     \end{minipage}
     \caption{\it
The varying of $v$, $d_1$, $d_2$, and $d_3$ with the
running scale $t$ ($t=\ln(m_0/m_W)$). The matching scale
is the mass of Higgs
scalar, which is taken to be $m_0=1200$ GeV.
The solid lines are the results of the RGE method,
while the dashed ones are those of the direct method.}
\label{fig3}
\end{figure}

In the RGE method, it becomes quite manifest that the effects
of heavy DOF to the low energy dynamics are related with two
factors, 1) the mass of the heavy particle, which determines
the matching scale $\mu_{UV}$, and 2) the initial
values of anomalous couplings at the matching scale
determined by integrating out the heavy particle,
which are related with the spin of the heavy particle
and the strength of its couplings to the low energy DOFs.
If a heavy field doesn't participate in the process of
symmetry breaking, it will not contribute to the anomalious couplings
up to the $O(p^4)$ order
and its effects can be estimated by the
decoupling theorem \cite{dcpl}.

As we know, there are several ways for the
$SU(2)$ breaking to its subgroups,
$SU(2)$ breaks to $U(1)$ \cite{georgi}, for instance.
In this paper, we only assume that the symmetry is broken from
a local one to a global one, where all components of vector boson
have the same mass. For the way of $SU(2)$ breaking to $U(1)$, the
effective Lagrangian will be more complicated. Several of the patterns of
symmetry breaking will be discussed in our next paper \cite{our2} when we consider
the renormalization of electroweak chiral Lagrangian.

Meanwhile, for the sake of simplicity,
no fermion field is taken into account, which
might introduce terms of anomaly.
Also, we have not included all of terms
breaking the charge, parity, and both symmetries.
If included,
the above procedure will be more complicated due to the
properties of the complete antisymmetric
tensor $\epsilon^{\mn\delta\gamma}$.
However, in principle, we can still make
the renormalization procedure order by order
even for the complicity.

The renormalization procedure in this paper
can easily be extended to study the
renormalization of the nonlinear sigma model
with $SU(N_f)$ symmetry \cite{gasser}, which has a very imporant
role to describe the low energy hadronic physics.
We will apply the related conceptions and methods to
the renormalization of the electroweak chiral effective
Lagrangian and the QCD chiral Lagrangian \cite{our2}.

\section{Acknowledgments}
One of the author, Q. S. Yan, would like to thank
Professor C. D. L\"u in the theory division of IHEP of CAS
for helpful discussion. And special thanks to
Prof. Y. P. Kuang and Prof. Q. Wang from physics department
of Tsinghua university,
for their kind help to ascertain some important points
and to improve the representation. To manipulate the algebraic
calculation and extract the $\beta$ functions of RGEs,
we have used the FeynCalc \cite{fcalc}.
The work of Q. S. Yan is supported by
the Chinese Postdoctoral Science Foundation
and the CAS K. C. Wong Postdoctoral Research Award Foundation.
The work of D. S. Du is supported by the
National Natural Science Foundation of China.

\end{document}